\documentclass[aps,twocolumn,prb,showpacs]{revtex4}
\usepackage{psfig}
\newcommand{\bea}{\begin{eqnarray}}
\newcommand{\beq}{\begin{equation}}
\newcommand{\eea}{\end{eqnarray}}
\newcommand{\eeq}{\end{equation}}
\begin{document}
\title{Spin splitting and precession in quantum dots with spin-orbit 
coupling: the role of spatial deformation}
\author{Manuel Val\'{\i}n-Rodr\'{\i}guez}
\affiliation{Departament de F{\'\i}sica, Universitat de les Illes Balears,
E-07122 Palma de Mallorca, Spain}
\author{Antonio Puente}
\affiliation{Departament de F{\'\i}sica, Universitat de les Illes Balears,
E-07122 Palma de Mallorca, Spain}
\author{Lloren\c{c} Serra}
\affiliation{Departament de F{\'\i}sica, Universitat de les Illes Balears,
E-07122 Palma de Mallorca, Spain}
\date{April 30, 2003}
\begin{abstract}
Extending a previous work on spin precession in GaAs/AlGaAs 
quantum dots with spin-orbit coupling, we study the role of 
deformation in the external confinement. Small elliptical deformations
are enough to alter the precessional characteristics at low magnetic
fields. We obtain approximate expressions for the modified $g$ factor
including weak Rashba and Dresselhaus spin-orbit terms. 
For more intense couplings numerical calculations are performed.
We also study the influence of the magnetic field orientation on the 
spin splitting and the related anisotropy of the $g$ factor.
Using realistic spin-orbit strengths our model calculations can 
reproduce the experimental spin-splittings
reported by Hanson et al.\ (cond-mat/0303139) for a one-electron dot.
For dots containing more electrons, Coulomb interaction effects are estimated 
within the local-spin-density approximation, showing that many features of the 
non-iteracting system are qualitatively preserved. 
\end{abstract}
\maketitle

\section{Introduction}

In the last years the study of spin-related phenomena has become one of 
the most 
active research branches in semiconductor physics.
The present advances in spin-based electronics\cite{Wol01} and 
the hope for better devices, with enhanced performance with respect 
to the conventional charge-based ones, encourage this research.
Two physical mechanisms underlie the operation of most {\em spintronic} 
devices:
a) the spin-spin interaction, present in ferromagnetic materials and in 
diluted magnetic semiconductors; and 
b) the electron spin-orbit (SO) coupling
stemming from relativistic corrections to the 
semiconductor Hamiltonian.
It should also be mentioned that, as shown recently by Ciorga {\em et al.}, 
another possibility of spin control involves the use of external 
magnetic fields to induce changes in the spin structure of a quantum 
dot.\cite{Cio03} 
These spin modifications affect the passage of currents through the system,
originating the spin blockade effect.    
A conspicuous example of device exploiting the SO coupling is
the spin transistor, first proposed by Datta and Das.\cite{Datxx} 
In this system the spin rotation induced by an adjustable Rashba 
coupling is used to manipulate the current. 

In a recent work,\cite{Valxx} we studied the spin precession of quantum dots
with SO coupling under the action of a vertical magnetic 
field of modulus $B$. It was shown that the SO coupling modifies
the precessional frequency from the Larmor expression 
$\hbar\omega_L=|g^*| \mu_B B$,
where $g^*$ is the bulk effective $g$ factor and $\mu_B$ is the 
Bohr magneton, to a different value depending on the dot 
quantum state. Namely, the modified precessional energy equals 
the gap between spin up and down states for the active 
level, the so called spin-flip gap $\Delta_{sf}$.

Purely circular dots are characterized by   
discontinuous jumps in angular momentum
with the number of electrons $N$ and the magnetic field, with a similar 
behavior for the precessional frequency. 
An interesting prediction of Ref.\ \onlinecite{Valxx} was that for some 
values of $N$ a finite $\Delta_{sf}$ persists even 
at $B=0$, i.e., a 
constant offset to the above Larmor formula. It is our aim in this work
to extend those investigations by including deformation in the external
confinement, as well as a more general treatment of the SO coupling,
considering Rashba and Dresselhaus contributions on an equal footing.
We shall show that small elliptical deformations are enough to sizeably
alter the precessional frequency, yielding a deformation-dependent 
$g$ factor and washing out the low $B$ offsets of purely circular dots.
Anisotropy effects in the $g$ factor will
also be studied by allowing for a tilted orientation of the magnetic field
vector with respect to the dot plane.

Spin dynamics in semiconductor nanostructures can be experimentally 
monitored with optical techniques. Indeed, a time-delayed laser 
interacting with a precessing spin experiences the Faraday rotation
of its polarization. Measuring the rotation angle for different delays 
allows to map the spin orientation and thus observe in detail
the dynamics. This technique has been applied to bulk semiconductors
(see Ref.\ \onlinecite{Kik99} for a recent review) and, also, to CdSe 
excitonic quantum dots in Ref.\ \onlinecite{Gup99}. Alternative methods
to gather information on the $g$ factor in quantum dots normally use
measurements of the resonant tunneling currents through the system
that permit the determination of the spin splittings and, therefore, deduce the 
effective $g$ value.\cite{Fuji02,Han03} 

Electron spin in quantum dots is much more stable than in bulk semiconductors,
due to the suppression of spin flip decoherence mechanisms.\cite{Kha00} 
Spin relaxation is predicted to occur on a time scale of 1 ms 
for $B=1$ T. Accordingly, in this work we shall neglect spin relaxation,
focussing on the much faster spin precession in quantum dots.
The spin splittings will be compared with those measured in 
Ref.\ \onlinecite{Han03},
showing that realistic values of the SO strengths can indeed reproduce the 
observed behavior.   
The paper is organized as follows. Section II presents the analytical
model for low SO intensities. In Sec.\ III we discuss the numerical results
for a variety of situations; namely, arbitrary SO strengths (A), 
tilted magnetic fields (B), one-electron dots (C) and treating 
Coulomb interaction effects (D) within
the local-spin-density approximation (LSDA). Finally, the conclusions are
presented in Sec.\ IV. 

\section{The Model}

\subsection{The noninteracting Hamiltonian}

Our model of a single quantum dot consists in $N$ electrons of effective
mass $m^*$  whose motion is restricted to the $xy$ plane where
an electrostatic potential $V_{\it ext}({\bf r})$ induces the confinement. 
We assume a GaAs host semiconductor, for which $m^*=0.067 m_e$.
To allow for elliptically deformed shapes we consider an anisotropic
parabola, i.e., 
\begin{equation} 
V_{\it ext}({\bf r})= \frac{1}{2}m^*(\omega_x^2 x^2 + \omega_y^2 y^2)\; .
\end{equation}
Neglecting for the moment Coulomb interactions between electrons we treat
the Hamiltonian for independent particles 
${\cal H}_{ip}=\sum_{i=1}^{N}{h(i)}$. 
The single-electron Hamiltonian ($h$) contains the kinetic/confinement
energy ($h_0$), the Rashba ($h_R$) and Dresselhaus ($h_D$) SO terms
and the Zeeman energy ($h_Z$);
\begin{equation}
h = h_0+h_R+h_D+h_Z\; .
\end{equation} 
The explicit expressions of $h_0$ and $h_Z$ read
\begin{eqnarray}
h_0 &=& \frac{{\bf P}^2}{2m^*}+V_{\it ext}(x,y)\; ,\\
h_Z &=& \frac{1}{2} g^* \mu_B (B_x\sigma_x+B_y\sigma_y+B_z\sigma_z)\; ;
\end{eqnarray}
where  
${\bf P}=-i\hbar\nabla+\frac{e}{c}{\bf A}$ represents the canonical 
momentum depending on the vector potential ${\bf A}=B_z/2(-y,x)$
and the $\sigma$'s are the Pauli matrices (used also in the SO contributions).
Note that all three components of the magnetic field contribute to the 
Zeeman term while only the vertical one couples with the kinetic 
energy through the vector potential. The GaAs bulk $g$ factor is 
$g^*=-0.44$.
Finally, the Rashba and Dresselhaus SO Hamiltonians may be written as\cite{Vosxx}
\begin{eqnarray}
\label{eqR}
h_{R} &=& \frac{\lambda_R}{\hbar}  
\left(\, P_y\sigma_x-P_x\sigma_y\,\right) \; ,\\
\label{eqD}
h_{D} &=& \frac{\lambda_D}{\hbar}  
\left(\, P_x\sigma_x-P_y\sigma_y\,\right) \; .
\end{eqnarray}  
The coupling constants $\lambda_R$ and $\lambda_D$ determine the SO strengths
and their actual values may depend on the sample. Several experiments on 
quantum wells have recently provided valuable information about realistic
ranges of variation for these coefficients.\cite{Nitxx} 

\subsection{The analytical solution}

It is possible to obtain analytical solutions 
when $h_0\gg (h_R\simeq h_D) \gg h_Z$ and $B_x=B_y=0$. In this case one
may use unitary transformations (as suggested in Ref.\ \onlinecite{Ale01})
yielding a diagonal transformed Hamiltonian. In a recent work\cite{Valyy} 
we used this technique to show that the SO (Dresselhaus) coupling
induces oscillations between up and down spin states when the magnetic 
field or the dot deformation are varied. Generalizing the 
transformations to consider both SO terms we define
\begin{eqnarray}
\tilde{h} &=& U_1^\dagger h U_1\; , \nonumber\\
U_1 &=& \exp\left\{\rule{0cm}{0.45cm}
-i\frac{m^*}{\hbar^2}\left[\rule{0cm}{0.35cm}
\lambda_R(y\sigma_x-x\sigma_y) \right.\right. \nonumber\\
&&  \qquad\qquad\quad +\left.\rule{0cm}{0.45cm} \left. \rule{0cm}{0.35cm}
\lambda_D(x\sigma_x-y\sigma_y)
\right]
\right\}\; .
\end{eqnarray}
Expanding in powers of the $\lambda$'s one finds for the transformed Hamiltonian
\begin{eqnarray}
\label{eqt}
\tilde{h} &=&  {{\bf P}^2\over 2m^*} 
+ V_{\it ext}(x,y) \nonumber\\
&-& (\lambda_R^2-\lambda_D^2) \frac{m^*}{\hbar^3}\, L_z \sigma_z 
+ \frac{1}{2}\, g^* \mu_B B_z \sigma_z \nonumber\\
&-& N (\lambda_R^2+\lambda_D^2) \frac{m^*}{\hbar^2}
+ O(\lambda^3)\; ,
\end{eqnarray}
where we have defined the {\em canonical} angular 
momentum operator $L_z=xP_y-yP_x$. Note that to $O(\lambda^2)$,
with $\lambda$ referring to both $\lambda_R$ and $\lambda_D$,
the Hamiltonian of Eq.\ (\ref{eqt}) is diagonal in spin space.
Nevertheless, the $x$ and $y$ spatial degrees of freedom
are still coupled through the vector potential in the kinetic 
energy and in $L_z$.

A second transformation for each spin subspace of Eq.\ (\ref{eqt}) 
may be used to obtain spatially decoupled oscillators. 
Namely, introducing a renormalized cyclotron frequency
\begin{equation}
\omega_{c\eta}={eB_z\over m^*c}+ (\lambda_D^2-\lambda_R^2)\, 
\frac{2m^*}{\hbar^3}\,s_\eta \; ,
\end{equation}
where $s_\eta=\pm 1$ for $\eta=\uparrow,\downarrow$, in Eqs.\ (5) of   
Ref.\ \onlinecite{Valyy} one obtains the masses $M_{k\eta}$ and frequencies 
$\Omega_{k\eta}$ of the two ($k=1,2$) decoupled oscillators for each spin.
Analogously, Eqs.\ (7) of that reference yields the eigenvalues 
$\varepsilon_{N_1N_2\eta}$, depending on the corresponding number of quanta 
in each oscillator ($N_1,N_2$). For completeness of the presentation we
repeat here the expressions for the latter two quantities,
\begin{eqnarray}
\label{eqm}
\Omega_{k\eta} &=& \frac{1}{\sqrt{2}}
\left(\rule{0cm}{0.5cm}\right.
\omega_x^2+\omega_y^2+\omega_{c\eta}^2\nonumber\\
& \pm & 
\sqrt{\left(\omega_x^2+\omega_y^2+\omega_{c\eta}^2\right)^2
-4\omega_x^2\omega_y^2}
\left.\rule{0cm}{0.5cm}\right)^{1/2}\; ,
\end{eqnarray}
where the upper (lower) sign in $\pm$ corresponds to $k=1$(2), and
\begin{eqnarray}
\label{spene}
\varepsilon_{N_1N_2\eta} &=&
\left( N_1+\frac{1}{2} \right) \hbar\Omega_{1\eta} +
\left( N_2+\frac{1}{2} \right) \hbar\Omega_{2\eta} \nonumber\\
&+& s_\eta \frac{1}{2} g^* \mu_B B_z \; ,
\end{eqnarray}

As a direct application of the above results we may write the effective
$g$ factor for precession around a vertical magnetic field from the
difference between the up and down single particle energies ($\Delta_{sf}$) 
with fixed oscillator quanta $N_1$ and $N_2$,
\begin{eqnarray}
\label{gfact}
|g| &\equiv& \frac{\Delta_{sf}}{\mu_B B}\nonumber\\
& \!\!\!\!\! =& 
\!\!\!\!\!
\left| g^* + \frac{\hbar}{\mu_BB_z}\sum_{k=1,2}{\left(N_k+\frac{1}{2}\right)
(\Omega_{k\uparrow}-\Omega_{k\downarrow})}\right|\; .
\end{eqnarray} 
This equation shows that in the general case the 
$g$ factor is actually a function of the electron state (through the quanta),
the SO coupling constants and the vertical magnetic 
field $B_z$ (through the $\Omega$'s). It is also worth to mention 
that since the energy gap $\Delta_{sf}$ and the modulus of the magnetic 
field ($B$) are positive quantities, only the absolute value of the $g$
factor is determined by Eq.\ (\ref{gfact}).

\begin{figure}[t]
\centerline{\psfig{figure=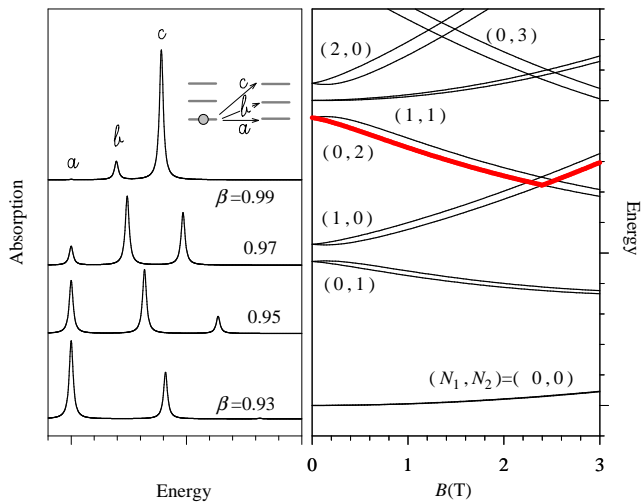,width=3.35in,clip=}}
\caption{Right panel: Evolution of the single particle energies as a 
function of vertical magnetic field. Each doublet corresponds to different spin
orientations in the transformed frame (Sec.\ II.B). 
The SO intensity is fixed at 
$\lambda_D^2-\lambda_R^2=(1.2\times10^{-9}\,{\rm eV}\,{\rm cm})^2$.
The level responsible for the spin-flip transition when $N=7$ is marked with 
a thick line.
Left panel: Strength of the spin-flip excitation for different deformations
($\omega_x=\beta\omega_y$; $\hbar(\omega_x+\omega_y)=12$ meV). The inset 
characterizes the transitions for $N=7$ and $B=0$ of the right panel, with 
$a$ indicating the transition between Kramers conjugates.}
\label{fig1}
\end{figure}

\subsection{The transition between Kramers conjugates}

When ${\bf B}$ vanishes the 
full Hamiltonian fulfills time-reversal symmetry and, according to a well known 
theorem of quantum mechanics, in that limit a degeneracy should prevail 
(Kramers degeneracy). As shown in Fig.\ 1, the single-particle energies 
$\varepsilon_{N_1N_2\eta}$ indeed 
merge into degenerate pairs at vanishing magnetic field. These pairs 
are split by the combined action of the SO and magnetic field contributions
and for a given $(N_1,N_2)$ one obtains parallel doublets when 
increasing $B$. Depending on the sign of $\lambda_D^2-\lambda_R^2$ the lower
member of each doublet will have a given spin orientation {\em in the 
transformed frame}; namely, upwards for positive sign and 
downwards for negative sign.

If the system has good angular momentum in the intrinsec reference frame ($L_z$), 
as happens in a circular confinement $\omega_x=\omega_y$, the Kramers 
conjugates at $B=0$ possess opposite angular momenta. Therefore, the  
spin flip transition between them is forbidden since the relevant 
matrix element preserves angular momentum. 
On the contrary, when the system
is deformed, for instance due to an anisotropic confinement
$\omega_x=\beta\omega_y$, 
the spin flip transition between Kramers conjugates becomes possible
since angular momentum is no longer a 'good' quantum number.
This key point determines qualitatively different spin 
precessional spectra. In fact, when the transition between conjugates is forbidden 
there is a gap in the spectrum and a non-vanishing precession frequency
at $B=0$ (the precessional offset discussed in Ref.\ \onlinecite{Valyy}). 
This gap vanishes if the transition between conjugates is allowed due to the 
deformation. 
In the left panel of Fig.\ 1 we show the evolution of the precessional
peaks as the deformation is reduced ($\beta\to 1$). As discussed, 
the transition between Kramers conjugates ($a$) switches off 
when approaching the circular case.

\begin{figure}[t]
\centerline{\psfig{figure=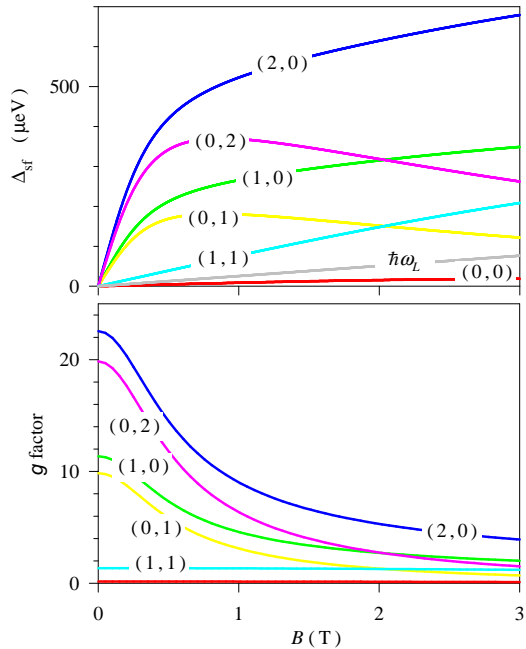,width=2.75in,clip=}}
\caption{Upper: spin-flip energy gap for the 
different levels of Fig.\ 1.
The Larmor energy $\hbar\omega_L$ is also indicated. 
Lower: g factors in absolute value
inferred from the upper panel results as 
$|g|=\Delta_{sf}/(\mu_B B)$.}
\label{fig2}
\end{figure}

\subsection{The $g$ factors}

The upper panel of Fig.\ 2 displays the spin-flip gap
for different levels, characterized
by their oscillator quanta in the transformed frame. 
The lower panel shows the corresponding $g$ factors obtained from 
$\Delta_{sf}$ and the modulus of the magnetic field using the first 
equality of Eq.\ (\ref{gfact}).
As in Fig.\ 1 a SO value of 
$\lambda_D^2-\lambda_R^2=(1.2\times10^{-9}\,{\rm eV}\,{\rm cm})^2$
as well as a deformation
of $\beta=0.9$ have been assumed. We note that there is a strong dependence 
of the precessional
properties on the electronic state, with many cases showing a dramatic 
deviation from the Larmor result. 
When the number of quanta is shared asymmetrically between the two
oscillators the $g$ factor takes very large values at small magnetic fields,
decreasing quite abruptly with $B$. On the contrary, when $N_1=N_2$ there 
is a rather flat $B$-dependence of the $g$ factor and lower
enhancements. Note also that spin-flip energies below the Larmor
result are obtained for the $(0,0)$ state, implying a $g$ factor lower
than the bare value. We have checked that other values of the SO couplings
and dot deformations do not lead to qualitative variations of this behaviors
although, obviously, the numerical values are changed.

\section{Cases of numerical treatment}

When the SO coupling can not be considered weak or when the magnetic field
points in a tilted orientation, with respect to the $z$ axis, the above 
analytical treatment does not remain valid. One must then resort to direct
numerical solution of the single-particle Schr\"odinger equation  
\begin{equation}
h\,\varphi_i({\bf r},\eta) = \varepsilon_i\, \varphi_i({\bf r},\eta)\; .  
\end{equation}
As in Ref.\ \onlinecite{Valyy},
we have proceeded by discretizing in a uniform grid of points, finding
the orbitals and energies $\{\varphi_i({\bf r},\eta),\varepsilon_i\}$
using matrix techniques. In terms of these results one
 can directly compute the spin-flip 
strength function,
\begin{eqnarray}
\label{eq10}
S_{\it prec}(\omega) &=& \sum_{ij}{
(1-f_i)f_j\, 
\left|\langle \varphi_i | \sigma_x | \varphi_j \rangle\right|^2} \nonumber \\
&\times& \delta(\varepsilon_i-\varepsilon_j-\hbar\omega) 
\; ,
\end{eqnarray}
where $i$ and $j$ span the whole single particle set and the $f_i$'s 
give the orbital occupations. 

\begin{figure}[t]
\centerline{\psfig{figure=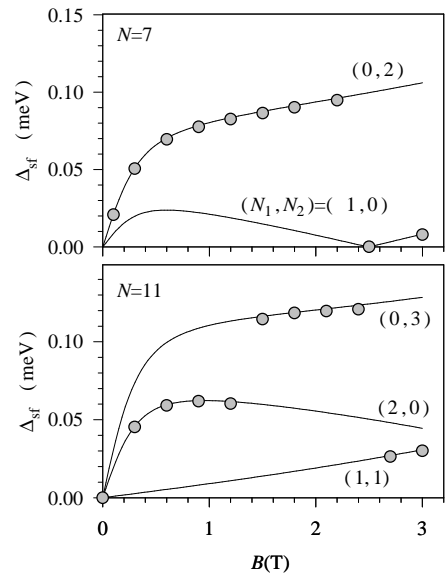,width=2.3in,clip=}}
\caption{ Numerical results for the spin-flip gap when $\lambda_R=0$
and $\lambda_D=0.5\times 10^{-9}$ eVcm. For comparison the solid line display the 
analytical results from Eq.\ (\ref{gfact}).}
\label{fig3}
\end{figure}
\begin{figure}[t]
\centerline{\psfig{figure=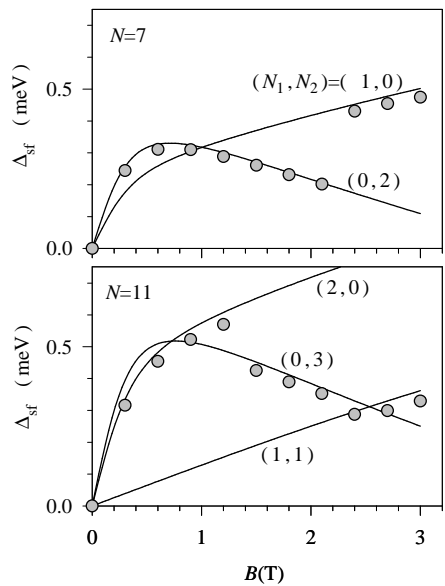,width=2.3in,clip=}}
\caption{Same as Fig.\ 3 but for $\lambda_R=1.2\times 10^{-9}$ eVcm 
and $\lambda_D=0$.}
\label{fig4}
\end{figure}

\subsection{Vertical magnetic fields}

We have checked that the numerical solution recovers the previously discussed 
analytical limit for vertical magnetic fields and weak SO couplings. For instance, 
Fig.\ 3  compares the spin-flip gaps   
for cases with a weak pure Dresselhaus coupling having $N=7$  and 11 
electrons. 
An excellent agreement between  the numerical data and the prediction of 
Eq.\ (\ref{spene}) is found. Note that in the numerical case discontinuous jumps 
in the evolution of  $\Delta_{sf}$ as a function of $B$ are  
obtained whenever the ground-state solution implies a reordering
of levels in energy. 
Figure 4 displays a similar result for a  pure Rashba
coupling, with a somewhat stronger intensity. Small deviations can be seen 
with the analytical result, although the agreement 
is still quite good. Our results thus indicate that 
the analytic treatment works rather well for SO couplings as large as
$1.2\times 10^{-9}$ eV$\,$cm, which is in the range of the experimentally achieved values.

\begin{figure}[t]
\centerline{\psfig{figure=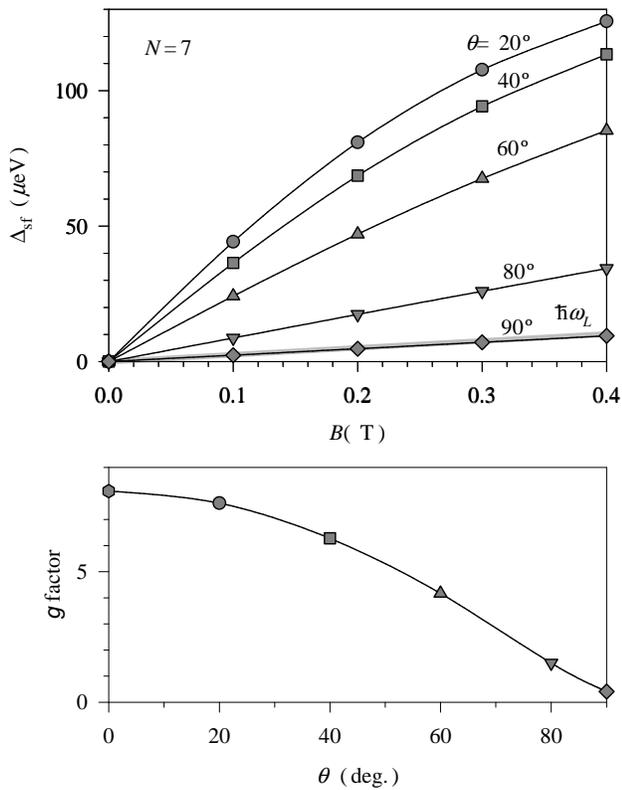,width=3.25in,clip=}}
\caption{Upper: Dependence of the spin-flip gap on the tilting angle 
of the magnetic field with respect to the vertical direction. The thick gray 
line shows the Larmor energy.
Lower: Variation of the g factors in the limit $B\to 0$ as a function of 
the tilting angle $\theta$.}
\label{fig5}
\end{figure}

\subsection{Tilted magnetic fields}

In Fig.\ 5 we have analyzed the dependence of the precessional properties on
the tilting angle of the magnetic field with respect to the $z$ axis,
zero angle meaning perpendicular magnetic field and $\theta=90^{\rm o}$ 
parallel ${\bf B}$ to the plane of motion. Note that the spin-orbit interaction 
is not invariant under rotations in the $x-y$ plane so that its effects
depend on the particular direction of tilting. In practice, however, 
different directions lead to only subtle differences, whilst the 
strong dependence is given by the angle $\theta$. For this reason we only 
discuss the case of tilting along the $x$-axis.
We find a rather strong dependence of the spin-flip gap on the tilting angle,
with a maximum deviation from the Larmor energy for perpendicular field.
When the tilting angle is increased a smooth energy decrease in the direction of
the Larmor value is seen. Actually, for parallel orientation the results are slightly
below the Larmor line.
In the lower panel of Fig.\ 5 the $g$ factors in the limit of vanishing magnetic field
are displayed. In correspondence with the transition energies
the largest deviations from the bulk value are obtained for the perpendicular
direction while the parallel $g$ factor is more similar to the bare factor (0.44).
These results can be understood by noting that the SO mechanism couples 
better with the ${\bf B}$-induced currents in the perpendicular geometry
and, therefore, a larger influence on the spin precession is expected in 
this case.

\begin{figure}[t]
\centerline{\psfig{figure=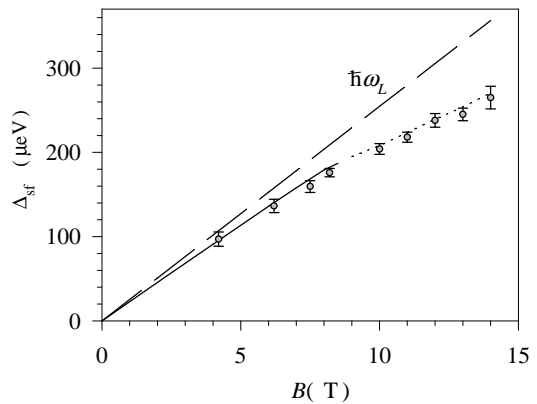,width=2.75in,clip=}}
\caption{ Experimental spin-flip energy gaps measured in Ref.\ \onlinecite{Han03}
for a one-electron dot. The solid line is the theoretical result
obtained using the experimentally known $\omega_0$ values while
the dotted extension is a fit (see Sec.\ III.C).}
\label{fig6}
\end{figure}

\subsection{A comparison with experiment}

In a recent experiment Hanson {\em et al.}\cite {Han03} have
measured the spin splitting in a one-electron dot by means of 
conductivity experiments using a parallel magnetic field. It is our 
purpose here to show that 
the SO-induced modifications can be the source of the observed
deviation of the spin-flip energy with respect to the Larmor
result. As stated in the previous section, when the magnetic field
is alligned parallel to the plane of electronic motion the spin splitting
recovers a Zeeman-like behavior with an effective $g$-factor
slightly smaller than the bulk value. This reduction of
the spin splitting is enhanced as the spacing of the orbital levels is
reduced, i.e., spin-orbit interaction induces a compression of the
spin levels as $\omega_x$ and $\omega_y$ become smaller.

In Fig.\ 6 we display the results obtained for a circular 1-electron
dot  (deformation has no significant influence on the spin-splitting of
the lowest energy state) with feasible values of SO coupling. Namely,
we assumed $\lambda_R=0.35\times 10^{-9}$ eVcm, in the range of experimental
values for GaAs,\cite{Vag98} and $\lambda_D=0.8\times 10^{-9}$ eVcm.
This latter parameter is obtained by assuming a 2DEG
width $z_0 \simeq 60$~{\AA} in 
the formula $\lambda_D=\gamma(\pi/z_0)^2$, where
$\gamma=27.5$ eV{\AA}$^3$ is the GaAs specific constant.\cite{Vosxx}
We still need to input the external confinement frequency 
$\hbar\omega_0$ before the calculation can be performed.
Using for this parameter the measured values of the orbital level spacing, lying
between 0.96 and 1.1 meV for the range from $B=0$ to 8 T,\cite{priv}
one obtains the solid line of Fig.\ 6. For higher values of 
the magnetic field experimental values are not available
and we have inferred the $\hbar\omega_0$ values in order to fit
the measured spin splittings. By assuming $\hbar\omega_0\approx 0.5-0.6$ meV 
we obtain the dashed line in Fig.\ 6. Overall, the agreement with 
the measurements is rather good and, though this is a certainly a simplified 
model, we believe it indicates that 
SO coupling plays an important role in explaining the measured spin gaps
in this system.

\begin{figure}[t]
\centerline{\psfig{figure=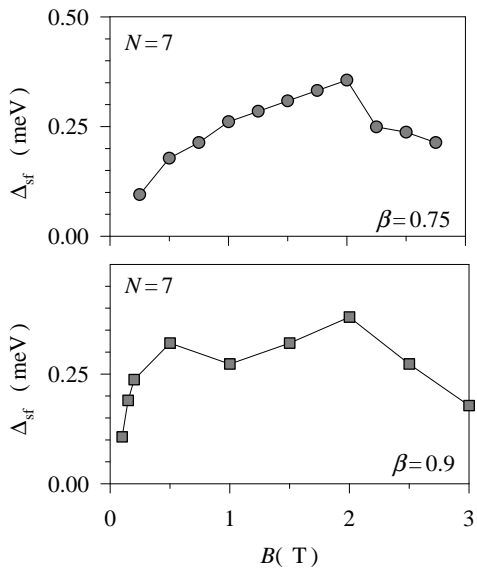,width=2.5in,clip=}}
\caption{Spin precessional energies within TDLSDA. The magnetic field points 
in the vertical ($z$) direction.}
\label{fig7}
\end{figure}

\subsection{Addition of Coulomb interactions}

The above sections have dealt with the SO-induced modifications of the 
spin precession in the absence of Coulomb interaction between electrons. 
We shall now estimate the role of the latter by resorting to the time-dependent
local-spin-density approximation (TDLSDA) for 
noncollinear spins. This approach was already used by us in Ref.\ \onlinecite{Valyy} 
for circular dots. The reader is addressed to that reference for more details on 
this formalism. Here we shall only mention that the integration in time of the 
TDLSDA equations allows us to monitor the spin precession and, in particular,
to extract the precessional frequencies. Since the selfconsistent 
parts of the mean-field potential are recomputed as the system evolves
in time one is effectively taking into account dynamical interaction effects. 
The formalism is thus equivalent to the random-phase approximation (well known
in many body theory) with an effective interaction.

Figure 7 shows for some representative cases the precessional frequencies in  
TDLSDA with SO coupling and deformation. A vertical magnetic field has been 
also included. 
We note that a qualitatively similar behaviour is found with respect to 
the preceding analytical results. 
In particular, we emphasize that at small $B$
the precessional frequency tends to vanish and that there are discontinuous 
jumps due to level rearrangements. 
It can be seen that, for a higher
deformation (smaller $\beta$), the $B$-dependence of the precessional frequencies 
is smoother, in agreement with the analytical model.
Comparing with the non-interacting results there is a sizeable modification
of the evolution in the low $B$ range. While in the non-interacting scheme 
we obtain $g$ factors of 20 and 6.2 for $\beta=0.9$ and 0.75, respectively, 
when interaction is included these values raise to 21.9 and 6.55.

\section{Conclusions}

In this work we have analyzed 
the role of the deformation in the confinement to determine, in conjunction with 
SO coupling and magnetic field, the spin precessional 
properties of GaAs quantum dots. At small magnetic 
fields the deformation closes 
the spin-flip energy gap by allowing the transition between Kramers
conjugate states. In practice, this implies that the precessional 
frequencies of deformed systems have no offsets at $B=0$. The associated 
$g$ factors depend strongly on the quantum dot electronic state and 
on the magnetic field direction. By tilting ${\bf B}$ from  
vertical to horizontal direction one may tune the $g$ factor from 
large values to results close to the bulk one. 

When the magnetic field points in the vertical direction and the SO coupling 
is weak an analytical treatment, yielding the spin-flip energies and 
$g$ factors is possible. This provides relevant insights for the analysis 
of other cases that can only be addressed with numerical approaches.
For the case of a one-electron dot in a horizontal magnetic field we have 
compared the results obtained with feasible values of the SO coupling 
constants with recent experiments. We believe this comparison indicates
that the SO coupling plays an important role in explaining the measured 
spin gaps in this system. For dots containing more electrons, the
role of the Coulomb interactions has been estimated within TDLSDA. 
Sizeable modifications of the single-particle picture have been obtained 
although, qualitatively, the main features are preserved.   
 
\begin{acknowledgments}
This work was supported by Grant No.\ BFM2002-03241 
from DGI (Spain). We thank R. Hanson for useful discussions and for providing 
us the data used in Fig.\ 6. 
\end{acknowledgments}

\end{document}